# PhoenixCloud: Provisioning Resources for Heterogeneous Cloud Workloads

Jianfeng Zhan, Lei Wang, Weisong Shi, Shimin Gong and Xiutao Zang

**Abstract**—As more and more service providers choose Cloud platforms, a resource provider needs to provision resources and supporting runtime environments (*REs*) for heterogeneous workloads in different scenarios. Previous work fails to resolve this issue in several ways: (1) it fails to pay attention to diverse RE requirements, and does not enable creating coordinated REs on demand; (2) few work investigates coordinated resource provisioning for heterogeneous workloads. In this paper, our contributions are three-fold: (1) we present an RE agreement that expresses diverse RE requirements, and build an innovative system *PhoenixCloud* that enables a resource provider to create REs on demand according to RE agreements; (2) we propose two coordinated resource provisioning solutions for heterogeneous workloads in two typical Cloud scenarios: first, a large organization operates a private Cloud for two heterogeneous workloads; second, a large organization or two service providers running heterogeneous workloads revert to a public Cloud; and (3) A comprehensive evaluation has been performed in experiments. For typical workload traces of parallel batch jobs and Web services, our experiments show that: a) In the first Cloud scenario, when the throughput is almost same like that of a dedicated cluster system, our solution decreases the configuration size of cluster by about 40%; b) in the second scenario, our solution decreases not only the total resource consumption, but also the peak resource consumption maximally to 31% with respect to that of EC2 + RightScale solution.

—————————— ◆ ——————————

## 1. INTRODUCTION

Traditionally, users tend to use a dedicated cluster system (DCS) to provide homogeneous services. The *runtime environment software (RE) that is responsible for managing cluster resources and workloads* plays an important role since it has great impact on resource utilization and quality of services of user applications. Traditional REs only support homogeneous workloads, for example, OpenPBS [30] for parallel batch jobs or Océano [1] for web services. The resource utilization rates of a DCS are varying. For unexpected peak loads, a DCS cannot provision enough resources, while lots of resources are idle for normal loads. Recently, several pioneer computing companies are adopting infrastructure as a service (IaaS). For example, as a resource provider, Amazon provides elastic computing cloud (EC2) services [8] to end users in order to offer outsourced resources in the granularity of XEN virtual machine [39]. A new term *Cloud* is used to describe this new computing paradigm [5] [33] [41]. We regard that the most appropriate one is defined in [38]. According to this definition, *a Cloud is a large pool of easily usable and accessible virtualized resources, which can be dynamically reconfigured to adjust to a variable load (scale), allowing also for optimum resource utilization.*

As more and more service providers choose Cloud platforms, a resource provider (which can be regarded as a Cloud infrastructure provider) needs to provision REs *for heterogeneous workloads in different scenarios*. For example, a large organization operates two DCSes for its two affiliated departments: a batch queuing system for parallel batch jobs for the first department and a Web service infrastructure for the second one. If this large organization wants to consolidate two heterogeneous workloads on a private Cloud or resort to a public Cloud, an enabling system needs to resolve two related issues: a) how does a resource provider create a RE on demand for different RE requirements? b) How does a resource provider provision resources when heterogeneous workloads are consolidated? Though a cloud system may imply geographically distributed systems [33], in this paper, when we refer to a cloud platform, we only consider it as a centralized cluster system (which is called as *a Cloud site* for clarity). A Cloud system can be a federated system of Cloud sites [33].

Previous work fails to resolve these issues in two ways. First, very few approaches pay attention to diverse RE requirements of service providers, including the large organization mentioned above, and no system enables creating coordinated REs on demand for heterogeneous workloads. *A coordinated RE is the one that can share coordinated resources with another RE*. For example, if the large organization chooses a Cloud platform, two REs belong to this condition. Most of previous efforts focus on service description languages for web service applications [25] or job definition languages for computational applications [15] or service definition mechanisms[12] for virtual execution environments. For example, in the recent work, based on the DMTF's Open Virtualization Format standard, F. Galán et al [12] propose a service specification language for cloud computing platforms in order to facilitate interoperability among IaaS clouds. They [12] [15] [25] are not qualified for describing diverse RE requirements in creating REs on demand. Besides, most of previous

————————————————
- Jianfeng Zhan, Lei Wang, Shimin Gong and Xiutao Zang are with Institute of Computing Technology, Chinese Academy of Sciences. E-mail: jfzhan@ncic.ac.cn.
- Weisong Shi is with Department of Computer Science, Wayne State University. E-mail:weisong@cs.wayne.edu.

efforts do not *treat RE as a first-class entity in the system design*, and can not provision REs on demand. In our opinion, *RE's being a first class entity* has three meanings: a) there is a RE agreement that is qualified for expressing diverse RE requirements; b) a RE can be created on demand according to a RE agreement; c) there is a framework that supports the development of a RE satisfying the new requirement. For example, D. Irwin et al [6] share the similar goal of our work by proposing a service oriented architecture prototype for resource providers and consumers to negotiate access to resources over time. However, their system *Shirako* does not explicitly support service providers to express customized RE requirements. R. S. Montero et al [29] propose an architecture to provision computing elements that focuses on resolving the growing heterogeneity (hardware and software configuration) of the organizations that join a Grid when porting a Grid application; however it does not focus on provisioning REs for heterogeneous workloads.

Second, few previous efforts investigate coordinated resource provisioning for heterogeneous workloads when they are consolidated on a Cloud platform. For example, M. Steinder et al [37] use high-level performance goals to drive resource allocation; however the proposed mechanisms in [37] only benefit a system with a *homogeneous, particularly non-interactive workload* by allowing more effective scheduling of jobs. Focusing on the specific problem of supporting workloads that combine advance reservation (resource) requests and best-effort (resource) requests, B. Sotomayor et al [36] present the design of lease management architecture, *Haizea* that implements leases as virtual machines (VMs) to provide leased resources with customized application environments. However, B. Sotomayor et al *[36] only consider homogeneous workloads (only parallel batch jobs) mixed with best-effort lease requests and advance reservation requests.*

In this paper, we design and implement an innovative system, PhoenixCloud, to facilitate a resource provider to provision REs on demand. The contributions of our paper are concluded as follows:

(1) We present a RE agreement that express diverse RE requirements and build an innovative system PhoenixCloud to enable creating REs on demand according to RE agreements.

(2) We propose two coordinated resource provisioning solutions for heterogeneous workloads in two typical Cloud scenarios: first, a large organization operates a private Cloud for two heterogeneous workloads (Web services and parallel batch jobs); second, a large organization or two service providers running heterogeneous workloads revert to a public Cloud.

(3) A comprehensive evaluation has been performed in experiments. For typical workload traces of parallel batch jobs and Web services, our experiments show that: a) in the first Cloud scenario, when the throughput is almost same like that of a DCS, our solution decreases the configuration size of cluster by about 40%; b) in the second Cloud scenario, our solution decreases not only the total resource consumption, but also the peak resource consumption maximally to 31% with respect to that of EC2 + RightScale solution.

This paper includes seven sections. Section 2 summaries the related work. Section 3 introduces several representative RE requirements. Section 4 explains PhoenixCloud design and implementation. Section 5 proposes two policies for coordinated resource provisioning. Section 6 evaluates our system, and Section 7 draws the conclusion.

## 2. RELATED WORK

In this section, we summarize related work of description models, enabling systems and resource provisioning.

## 2.1. Description models and systems

Most of previous efforts focus on service description languages for web service applications [25] or job definition languages for computational applications [15] or service definition mechanisms for virtualized execution environments [12][22][23][24][34]. EC2 allows end users to describe their resource requirements, e.g., virtual machines, and the EC2 extended service – RightScale [32] allows service providers to describe their requirements for Web services; A. Keller et al [25] propose a framework to specify service-level agreements for web services; A. Hoheisel et al [15] present a framework to define both workflow and dataflow for job applications. F. Galán et al [12] propose a service specification language for cloud computing platforms in order to facilitate interoperability among IaaS clouds, and also address important issues such as custom automatic elasticity and performance monitoring. R. Buyya et al[38] propose the meta-negotiation document to determine definition and measurement of user QoS parameters. However, they are not qualified for describing diverse RE requirements in creating REs on demand.

No previous efforts treat RE as a first-class entity in system design, and they provision either resources directly to end users [8] or *hosted application environments* without paying attention to different RE requirements of heterogeneous workloads. *Hosted application environment* [12] often consists of a collection of virtual machines (VM) with several configuration parameters for software components included in the VMs. EC2 directly provisions resources to end users. Without enabling the user role of service provider, EC2 relies upon end user's manual management of resources. EC2 extended services: RightScale [32], Enomalism [9] and GoGrid [13] systems provide automated cloud computing management systems that assist you in creating and deploying *only scalable Web service applications* running on EC2 platforms. D. Irwin et al [6] share the similar goal of our work by providing a *Shirako* prototype of service oriented architecture for resource providers and consumers to negotiate access to resources over time;

however *Shirako* does not explicitly support service providers to express personalized RE requirements.

M. Steinder et al [37] show that virtual machine allows heterogeneous workloads to be collocated on any server machine, and proposes the system architecture of managing heterogeneous workload. However, it does not treat RE as a first-class entity in the design. B. Rochwerger et al [27] pay attention to implement an architecture that would enable providers of cloud infrastructure to dynamically partner with each other. Their system *Reservoir* does not consider how to provision REs on demand on a Cloud site. R. S. Montero et al [29] propose an architecture to provision computing elements that focuses on resolving the growing heterogeneity (hardware and software configuration) of the organizations that join a Grid. A. Bavier et al [4] demonstrate dynamic instantiation of distributed virtualization in a wide-area testbed deployment with a sizable user base, whereby each service runs in an isolated slice of PlanetLab's global resources.

## 2.2. Resource provisioning

Few previous efforts discuss coordinated resource provisioning for heterogeneous workloads.

L. Grit et al [14] design a *Winks* scheduler to support a weighted fair sharing model for a virtual "cloud" computing utility, such as Amazon's EC2, where each request is for a lease of some specified duration for one or more virtual machines. The goal of the *Winks* algorithm is to satisfy these requests from a resources pool in a way that preserves the fairness across flows, while our work focuses on how to provision resources for heterogeneous workloads when they are consolidated on a Cloud site.

M. Steinder et al [37] only exploits a range of new automation mechanisms that will benefit a system with a *homogeneous, particularly non-interactive workload* by allowing more effective scheduling of jobs. By considering a workload in which massively parallel tasks that require large resources pools are interleaved with short tasks that require fast response but consume fewer resources, M. Silberstein et al [35] devise a scheduling algorithm. In nature, they only consider the parallel batch jobs with different resource demands. M.W. Margo et al [28] are interested in metascheduling capabilities (co-scheduling for Grid applications) in the *TeraGrid* system, including user-settable reservations among distributed cluster sites.

B. Lin et al provide an OS scheduling technique, *VSched* [26], for heterogeneous workload VMs. VSched enforces compute rate and interactivity goals for interactive workloads, including web workloads and non-interactive ones. It provides soft real-time guarantees for VMs hosted on a single server machine. B. Sotomayor et al [36] present the design of lease management architecture, *Haizea* th*at implements leases as virtual machines (VMs). VSched* and *Haizea* can be used as a component of our system for specific workloads.

## 3. DIVERSE RE REQUIREMENTS

In this section, we summarize several representative cases for discussing RE requirements on a Cloud site.

Case One: Some universities are trying outsourcing of HPC services, just taking in this way the role of job-execution service providers [3].

Case Two: many small companies have reverted to hosting environments for deploying Web services so as to decrease cost.

Case Three: a large organization has two representative departments: a batch queuing system for parallel batch jobs for the first department and a Web service infrastructure for the second one. Instead of operating two DCSes, the organization wants to consolidate heterogeneous workloads on a private Cloud or resorts to a public Cloud.

Three observations can be derived from the above three cases:

(1) There are three main user roles in the observed systems: a *resource provider*, *service providers* and *end users*. For example, in Case two, universities play the role of service providers, and they want to outsource resources to a resource provider and run batch queue systems for end users -graduate students or researchers.

(2) A resource provider *does need to provision REs for heterogeneous workloads*. For example, when the organization in Case Three chooses a private Cloud or resorts to a public Cloud, or two service providers in Case one and Case two resort to a public Cloud, a resource provider requires provisioning two different REs for heterogeneous workloads.

(3) For heterogeneous workloads, RE requirements are dramatically different. Coordinated resource provisioning for heterogeneous workloads may bring benefits to service providers and resource providers.

For example, REs for parallel batch jobs and Web services differ in four ways:

- Workloads are different. Web service workloads are often composed of a series of requests; while parallel batch job workloads are composed of a series of submitted jobs, and each job is a parallel or serial application.
- Resource consumptions are different. Running a parallel application needs a group of exclusive resources. While for Web services, requests will be serviced simultaneously and interleavedly through multiplex use of resources.
- Performance goals are different. From perspectives of end users, for parallel batch jobs, in general submitted jobs can be queued when resources are not available. However, for Web services like Web servers or search engines, each individual request needs an immediate response.
- Time scales of management are different [37]. Due to the nature of their performance goals and short duration of individual requests, Web services need automation at short control cycles, e.g., seconds; However, parallel batch jobs typically require calculation of a schedule for an extended period of time [37], e.g., hours.

When web service applications and parallel batch jobs are consolidated, we can create two *coordinated REs* and propose *coordinated resource provisioning* for two coordinated REs since they have different performance goals.

## 4. PHOENIXCLOUD DESIGN AND IMPLEMENTATION

In Section 4.1, we introduce the objectives of PhoenixCloud. Section 4.2 proposes the RE agreement. In Section 4.3, we describe the architecture.

### 4.1. Objectives

PhoenixCloud has several objectives:
1) Responsibility division between a resource provider and service providers. In our system, a resource provider is responsible for creating, destroying REs and provisioning resources to different REs on *a Cloud site,* while a service provider only focuses on providing service.
2) Provisioning a RE on a basis of a RE agreement. PhoenixCloud provides a RE agreement for a service provider to express RE requirements. According to a RE agreement, a RE is provisioned on demand.
3) Pluggable resources type [21]. Similar to *Shirako*, provisioned resources will include servers, storages, and network resources. Presently, our system mainly facilitates provisioning servers in the granularity of node or virtual machine.
4) Coordinated resource provisioning for two coordinated REs. If allowed by service providers, PhoenixCloud supports coordinated resource provisioning for two heterogeneous workloads.

PhoenixCloud evolves from our previous Phoenix system [40]. We have implemented PhoenixCloud on the Dawning 5000 cluster system, which is ranked as top 10 of Top 500 super computers in November, 2008. It is expected that PhoenixCloud will be deployed on the super computer-Dawning 6000 system in Shenzhen super computing center, China, in 2010.

PhoenixCloud has two major features: a) allows a service provider to express RE requirements and provisions a RE on demand according to a RE agreement; b) proposes *coordinated resource provisioning* for heterogeneous workloads.

### 4.2. RE Agreement

We present an RE agreement as a basis for provisioning a RE or two coordinated REs on demand.

In our opinion, in addition to *service-level agreements* between service providers and end users, *job definitions* for computational applications and *service definitions* for web services, both a resource provider and a service provider need a RE agreement to express diverse RE requirements, on a basis of which, a resource provider can flexibly provision REs on demand for service providers. Figure 1 shows the relationships of different agreements among different roles.

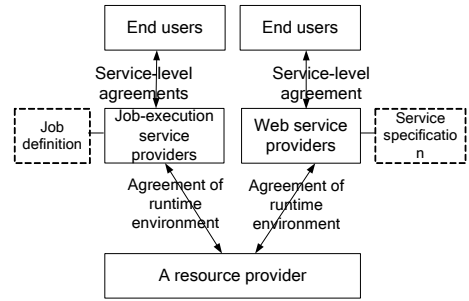

**Fig.1. The agreements among a resource provider, service providers and end users.**

A RE agreement includes the following information:
(1) Relationships between a service provider and a resource provider.

We support three different relationships: *same* or *affiliated* or *business*. The *same* relationship means that a single user plays the roles of both resource provider and service provider, which describes a DCS; the *affiliated* relationship means that a user playing the role of service provider is affiliated to a user playing the role of resource provider, which describes Case Three in Section 3; the *business* relationship means that a service provider has the *business* relationship with a resource provider, which describes Case One or Case Two in Section 3.

(2) Workload types.

Presently, we support two workloads types: *parallel batch jobs* and *Web services*.

(3) The allocation granularity of resources.

We support resource allocation in the granularity of nodes or virtual machines like XEN. For virtual machines, we provide predefined or user-defined virtual machine templates. For both nodes and virtual machines, users need to specify the customized operating system types and versions.

(4) Coordinated REs.

A service provider needs to decide two conditions: (a) whether a new RE has a coordinated RE that belongs to the same service provider; (b) Whether a service provider agrees that a new RE is coordinated to share resources with other RE of another service provider.

(5) Resource coordination models and bound sizes of resources.

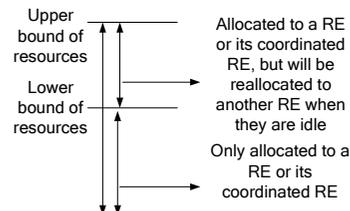

**Fig. 2. Two bound sizes of resources.**

In each RE, a service provider needs to specify two optional bound of resources: *the lower bound* and *the upper bound*. The *lower bound* is *rigid* in that a resource provider will guarantee that resources within the limit of *lower bound* will only be allocated to a RE or its coordinated RE. The *upper bound* is *flexible* in that resources of the range defined by *the lower bound* and *upper bound,* which *firstly* satisfy resource requests of *the* *specified RE* or its *coordinated REs*, can be reallocated to another *RE* when they are idle. Fig.2 shows the relationships between two bound sizes of resources.

For two typical heterogeneous workloads: Web services and parallel batch jobs, we propose *resource coordination models in two Cloud scenarios*.

In this first Cloud scenario, we presume that a resource provider owns the fixed resources in a private Cloud that satisfy the resource requests of two coordinated REs. For a RE, the sizes of lower bound and upper bound are same. For two coordinated REs, the size of *coordinated resources* that are shared by two REs is the sum of the lower bounds of two REs. We call this model the *FB* (Fixed Bounds) model.

In the second Cloud scenario, we presume that a resource provider owns enough resources that can satisfy the resource requests of N service providers (N>>2). For a RE, we only specify the size of the lower bound with the upper bound undefined. Each RE can request more resources beyond the limit of the lower bound. For two coordinated REs, the size of *coordinated resources* is the sum of *the lower bounds* of two REs. We call this model the *FLB_NUB* (Fixed Lower Bound and No Upper Bound) model.

(6) The setup policy.

The service provider determines when and how to perform the setup work when resources are dynamically requested or released. The setup work includes provisioning operating systems and configuring applications. For example, if the service provider pays high attention to the security of data, it may require wiping off the operating system and data on disks when a node is released to the resource provider.

Fig. 3 gives out the part of a RE agreement of parallel batch jobs for Case Three in Section 2. Our RE agreement is easily extensible, since we choose the XML (eXtensbile Markup Language) language to express it.

```
----------------------------------------------------------------------
<RE_agreement>
<relationship=" business "></relationship>
<type=" parallel_batch_jobs "></type>
<coordinated_RE="Yes">
</coordinated_RE>
<granularity="node "></granularity>
<resource_coordination_model="FLB_NUB"></resource_coordination_model>
<lower_bound_size="100"></lower_bound_size>
<upper_bound_size=null></upper_bound_size>
<setup_policy="NOOP"></ setup_policy >
</RE_agreement>
----------------------------------------------------------------------
```
**Fig.3. A part of a RE agreement.**

### 4.3. PhoenixCloud architecture

**Layered architecture**: PhoenixCloud follows a two-layered architecture: one is the *common service framework* (in short *CSF*) for a resource provider, and another is the *thin runtime environment software* (in short, *TRE*) for a service provider. The two-layered architecture has two implications: first, there lies a separation between the CSF and a TRE. The CSF is provided and managed by a resource provider, independent of any TRE. With the support of the CSF, a TRE or two coordinated TREs can be created on demand for a service provider. Second, for heterogeneous workloads, the common sets of functions of REs are delegated to the CSF, while a TRE only implements the core functions for a specific workload.

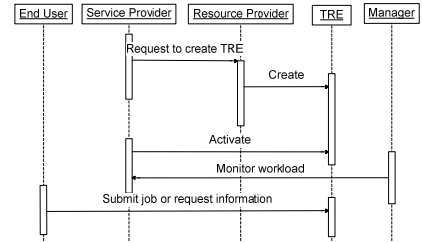

**Fig. 4. The interactions of three user roles.**

As shown in Fig.4, there are three interacting user roles in PhoenixCloud: a resource provider, service providers and end users:

- The CSF is running on the Cloud site. A resource provider is responsible for provisioning REs with the support of the CSF.
- The CSF provides *a Web portal* for a service provider to describe its RE requirements. After a service provider has requested to create a RE, the CSF is responsible for deploying and starting a TRE.
- After a service provider has activated its RE, a service provider has an associated *manager* that monitors workload changes and resources status. *The manager* is a core component of a TRE. *Each manager* requests or releases resources on behalf of the service provider according to load status and resources status.
- After a RE is providing service, end users use *the Web portal* to submit jobs or send requests.

The advantages of separating the CSF and a TRE have two points: first, developing a new TRE for different workloads is lightweight, since many common functions have been implemented in the CSF. Secondly, creating a TRE on demand is lightweight, since the CSF is ready and running before any TRE is created.

**Main components of** *the CSF*: The major components of the CSF are as follows:

(1) *The lifecycle management service* is responsible for managing the lifecycle of a TRE.

(2) *The resource provision service* is responsible for provisioning resources to a TRE.

(3) *The virtual machine provision service* is responsible for managing the lifecycle of a virtual machine, such as creating or destroying virtual machine, like XEN.

(4) *The deployment service* is a collection of services for deploying and booting the operating system, the CSF and TREs. Major services include DHCP, TFTP, and FTP.

(5) *The agent* on each node is responsible for discovering node resources, such as CPU information, memory size and operating system version;

downloading the required software package; starting or stopping service daemon, and transferring data.

(6) There are two types of *monitors*: *resource monitor* and *application monitor*. *Resource monitor* on each node monitors usages of physical resources, e.g. CPU, memory, swap, disk I/O and network I/O; *application monitor* monitors application status.

(7) *The process management service* is responsible for starting, signaling, killing, and monitoring parallel/sequential jobs.

**Main components of *a TRE*:** There are three components in a TRE: *the manager*, *the scheduler* and *the Web portal*. *The manager* is responsible for dealing with users' requests, managing resources and interacting with the CSF. *The scheduler* is responsible for scheduling the user's job or distributing user requests. *The web portal* is the GUI through which end users submit and monitor jobs or applications. When a TRE is created, a configuration file will describe their dependencies. The detail can be found in Section 4 of our previous work [43].

**The customized policies of the CSF and a TRE:** Fig.5 shows the major components and their extension points for the management mechanisms.

Specified for *the resource provision service*, *a resource provision policy* determines when *the resource provision service* provisions how many resources to *a TRE* or how to coordinate resources between *two coordinated REs*; *the setup policy* determines when and how to do the setup work, such as wiping off the operating system or doing nothing.

Specified for *the manager*, *the resource management policy* determines when *the manager* requests or releases how many resources from or to *the resource provision service* according to what policy.

For different workloads, *the scheduling policy* has different implications. For parallel batch jobs, *the scheduling policy* determines when and how *the scheduler* chooses parallel jobs for running. For Web service, *the scheduling policy* includes two policies: *the instance adjustment policy* and *the request distribution policy*. The *instance adjustment policy* decides when the number of Web service instances is adjusted to what an extent, and *the request distribution policy* decides how to distribute requests according to what criteria.

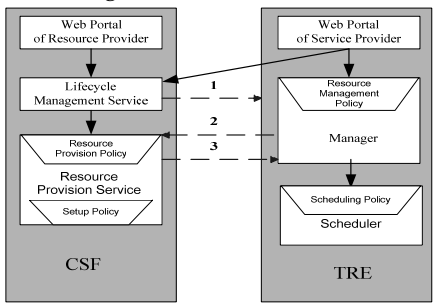

**Fig.5. The summary of interactions and extension points for the management mechanism of PhoenixCloud. Number 1 indicates creating, destroying, activating and deactivating a TRE; Number 2 indicates requesting and releasing resources; Number 3 indicates proactively provisioning resources.**

**Interactions of a TRE with the CSF:** In the rest of this paper, we call a TRE for parallel batch jobs as PBJ TRE; we call a TRE for Web service as WS TRE. Fig.6 shows the interactions between TREs and the CSF in two coordinated REs.

The interaction of a *WS TRE* with *the CSF* is explained as follows:

(1) *The WS manager* obtains resources with the size of *the lower bound* from *the resource provision service*, and runs the Web service instances with the matching scale.

(2) *The WS manager* interacts with *the load balancer* to set its *request distribution policy*. *The WS manager* registers the IP and port information of Web service instances to *the load balancer* that is responsible for assigning workload to Web service instances, and *the load balancer* distributes requests to Web services instances according to *the request distribution policy*. We integrate LVS [13] as the IP-level load balancer.

(3) *The monitor* on each node periodically checks the resources utilization rates and reports to the *WS manager*. If the threshold performance value is exceeded, e.g., *the average of utilization rates of CPUs consumed by instances exceeds 80%*, *the WS manager* adjusts the number of Web service instances according to *the instance adjustment policy*.

(4) According to current Web service instances, *the WS manager* requests or releases resources from or to *the resource provision service*.

The interactions of a PBJ TRE with the CSF are explained as follows:

(1) *The scheduling events* tell *the PBJ manager* to send scheduling command to *the scheduler*. *The scheduling events* include the timer registered by the administrator and new job arrival.

(2) *The scheduler* requests jobs and nodes information from *the PBJ manager*, and takes the decision to run jobs according to *the scheduling policy*.

(3) Driven by the periodical timer, *the PBJ manager* scans the jobs in queue. If the threshold values defined in *the resource management policy* are exceeded, *the manager* will request or release resources from or to *the resource provision service*.

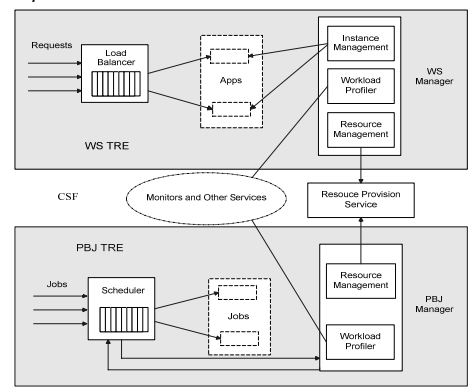

**Fig.6. Interactions of a PBJ TRE and a WS TRE with the CSF.**

**The lifecycle management of TREs:** A traditional RE is self-contained. PhoenixCloud facilitates creating a TRE on demand. Each TRE has three states: *uninitialized*, *created and running*. The *uninitialized* state indicates the nascent state of a TRE. The *created* state implies that a TRE for the specific workload is configured and deployed on a Cloud site. The *running* state indicates two meanings: first, resources with *the size of the lower bound* are allocated to a TRE; secondly, a TRE is providing service to end users.

By taking a RE agreement in Fig. 3 as an example, we introduce the major interactions as follows:

(1) Through *the Web portal of the resource provider*, a service provider creates its account, and then defines its RE requirements.

(2) Through *the Web portal of the resource provider*, the service provider sends *the message of creating RE* to *the lifecycle management service*. Then *the lifecycle management service* marks the state of the new RE as *uninitialized*.

(3) *The lifecycle management service* sends *the message of deploying RE* to *agents* on the related nodes, which requests *the deployment service* to download the required software package of the new TRE. After the new TRE is deployed, *the lifecycle management service* marks its state as *created*.

(4) The service provider sends *the message of activating RE* to *the lifecycle management service* through *the Web portal of the resource provider*.

(5) *The lifecycle management service* sends the configuration information of the new TRE to *the resource provision service*, including *the lower bound* and *upper bound* of resources, *the resource provision model*, *the setup policy*. For a new PBJ TRE, *the resource provision service* will search a WS TRE for coordinated resource provisioning if the service provider does not specify it.

(6) *The lifecycle management service* sends *the message of starting components of the new TRE* (which includes *the manager*, *the scheduler* and *the Web portal*) to *agents*. When the components of the new TRE are started, the command parameters will tell the components what policies should be taken. Then *the lifecycle management service* marks the state of the new TRE as *running*.

(7) Before resources are provisioned to the new *TRE*, *the setup policy* is triggered by *the resource provision service*. When the setup work is performed, *the resource provision service* notifies *the manager* that resources are ready.

(8) The new TRE begins providing service to end users.

(9) According to load status, *the manager* dynamically requests or releases resources, which will also trigger *the setup policy*.

To save the space, we omit the processes of deactivating and destroying a TRE.

**The advantage of PhoniexCloud:** The advantages of our system have two points: first, our system facilitates a service provider to express diverse *RE* requirements, and enables creating *REs* on demand. With *the RE agreement* as a basis, our system can adapt to different cases without the architecture change. For example, our system can adapt to three cases in Section 3. Second, our system supports coordinated resource provisioning for heterogeneous workloads, and our experiments in Section 6 show the benefit

# 5. RESOURCE COORDINATION AND MANAGEMENT POLICIES

In this section, we respectively propose policies for FB and FLB-NUB models in consolidating *two typical heterogeneous workloads: Web services and parallel batch jobs*.

## 5.1. The FB policy

We propose *the FB resource coordination policy* as follows:

(1) In creating two coordinated REs (a PBJ TRE and a WS TRE) for two heterogeneous workloads, service providers specify the same value for the *lower_bound_size* and *the upper_bound_size* for each RE.

(2) *The resource provision service* allocates resources with the sizes of the *lower bounds* to *two TREs* at their startups. *The size of coordinated resources* that are shared by *two coordinated REs* is the sum of *lower_bound_size* of two *REs*.

(3) Resource demands of *the WS TRE* have high priority than that of *the PBJ TRE*. If *the WS TRE* demands resources that can not be satisfied by *the resource provision service*, the latter will force *the PBJ TRE* to release resources with the size required by *the WS TRE*, and then reallocate resources to *the WS TRE*.

(4) *The resource provision service* registers *a periodical timer (a time unit of leasing resources)* for checking idle resources within the limit of *the size of coordinated resources* per *time unit of leasing resources*. If *there are idle resources*, *the resource provision service* will provision all idle resources to *the PBJ TRE*.

For the above resource provision policy, the matched *resource management policy* of *the PBJ TRE* is as follows:

(1) *The PBJ manager* receives the resources provisioned by *the resource provision service*.

(2) If *the resource provision service* forces *the PBJ manager* to return resources, the latter will release resources with the required size. If there are no enough idle resources in *the PBJ manager*, it will kill jobs from the beginning of the minimum job size in turn and release resources with the required size. If there are more than one running jobs with the same job size, the job with the latest starting time will be killed firstly.

In the rest of this paper, we call the above policies as *FB policies*.

## 5.2. The FLB-NUB policy

We propose *the FLB-NUB resource coordination policy* as follows:

(1) In creating two coordinated REs, service providers only specify the *lower_bound_size* for *each RE* with the *upper_bound_size* undefined.

(2) *The resource provision service* allocates resources with the sizes of *lower bound* to *the PBJ TRE* and *the WS TRE* at their startups.

(3) *The resource provision service* registers *a periodical timer (a time unit of leasing resources)* for checking idle resources within the limit of *the size of coordinated resources* per *time unit of leasing resources*. If *there are idle resources, the resource provision service* will provision all idle resources to *the PBJ TRE*.

(4) If *the WS TRE* demands resources, *the resource*

*provision service* will allocate enough resources.

For the above resource provision policy, the matched *resource management policy* of *a PBJ TRE* is as follows:

We define *the ratio of adjusting resource* as the ratio of *the accumulated resource demands of all jobs in queue* to *the current resources owned by a TRE*. When *the ratio of adjusting resource* is *greater than* one, it indicates that for immediate running, some jobs in the queue need more resources than that currently owned by a TRE.

We set *two threshold values of adjusting resources*, and respectively call them *the threshold ratio of requesting resource* and *the threshold ratio of releasing resource*.

The process of requesting and releasing resource are as follows:

(1) *The PBJ manager* registers *a periodical timer (a time unit of leasing resources) for adjusting resources per time unit of leasing resources*. Driven by the periodical timer, *the PBJ manager* scans jobs in queue.

(2) If *the ratio of adjusting resources* exceeds *the threshold ratio of requesting resource*, *the PBJ manager* will request resources with the size of *DR1* as follows:

DR1=the accumulated resources demand of all jobs in the queue –the current resources owned by *a PBJ TRE*.

(3) If *the ratio of adjusting resource* does not exceed *the threshold ratio of requesting resources*, but the ratio of *the resource demand of the present biggest job in queue* to *the current resources owned by a TRE* is *greater than one*, *the PBJ manager* will request resources with the size of *DR2*:

DR2= resources needed by the present biggest job in queue– the current idle resources owned by *a TRE*.

When the ratio of *the resources demand of the present biggest job in the queue* to *the current resources owned by a TRE* is *greater than one*, it implies that the largest job will not run without available resources.

(4) If *the ratio of adjusting resources* is lower than *the threshold ratio of releasing resources*, *the PBJ manager* will releases idle resources with the size of *RSS (ReleaSing Size)*.

RSS= the elastic factor * (idle resources owned by PBJ TRE), where 0 < *the elastic factor* < 1.

(5) If *the resource provision service* proactively provisions resources to *the PBJ manager*, the latter will receive resources.

In the rest of this paper, we call the above policies as *NLB-NUB policies*.

In a recent work of USENIEX 09 ATC, W. Zhang et al [42] argue that in managing *web services of data centers*, *actual experiments are cheaper, simpler, and more accurate than models* for many management tasks. We also hold the same position. In Section 6.4, we will explain how to obtain the management policies for a specific web service through real experiments.

# 6. PERFORMANCE EVALUATIONS

In this section, for Web services and parallel batch jobs, we compare the performance of PhoenixCloud, DCS and EC2+RightScale.

## 6.1. Evaluation metrics

For parallel batch jobs, the metrics are as follows: we choose the well known metrics- *the number of completed jobs* [3] [11] to reflect the major concern of a service provider. We use *the average turnaround time per jobs* to measure the main concern of end users. *The average turnaround time of jobs* is the time from submitting a job till completing it, averaged over all jobs submitted [11] [20].

For Web service, the metrics are as follows: we choose the well-know metrics, *throughput in terms of requests per second* to reflect the major concern of a service provider [6] [10]. For end users, we choose *the average response time per requests* to measure the quality of service, which reflects the major concern of end users [6] [10].

For two consolidated workloads, we choose *the total resource consumption* in terms of *node * hour* to evaluate the effectiveness of coordinated resource provisioning. We specially care about *the peak resource consumption* that is *the peak value of the resource consumption in terms of nodes*, since it is a key factor in the capacity planning of the system for a resource provider. *For the same workload, if the peak resource consumption of a system is higher, the capacity planning of a system is more difficult.*

We use *the accumulated times of adjusting resources* to evaluate *the management overhead* of a system, since *each event of requesting, releasing or provisioning resources* will trigger a setup action, for example wiping off the operating system or data. *The accumulated times of adjusting resources* are the times of resources being dynamically requested, released or provisioned when *a RE* is providing services.

All performance metrics are obtained in the same period that is the duration of workload traces.

## 6.2 Workload traces

(1) The workload traces of parallel batch jobs

We choose two typical workload traces from [31]. The utilization rate of all traces in [31] varies from 24.4% to 86.5%. We choose one trace with lower load-NASA iPSC trace (46.6% utilization) and one trace with higher load-SDSC BLUE trace (76.2% utilization).

*NASA iPSC* is a real trace segment of two weeks from Oct 01 00:00:03 PDT 1993. For NASA iPSC trace, the configuration of the cluster system is 128 nodes. SDSC BLUE is a real trace segment of two weeks from Apr 25 15:00:03 PDT 2000. For SDSC BLUE trace, the cluster configuration is 144 nodes.

(2) Web service workload

For Web service, we choose a real workload trace, *the World Cup workload trace* [2] from June 7 to June 20 in 1998. The World Cup workload is widely used in the research of resource provisioning for Web service applications, since it reflects the nature of a web service workload, of which *the ratio of peak load to normal load is high*.

## 6.3 Experiment methods

To evaluate and compare the DCS system, PhoenixCloud, and EC2+RightScale, we adopt the following experiments methods.

*a) The real experiments of World Cup workload*

For web service, we obtain *the resource consumption trace* through *the real experiments that deploys a WS TRE for*

*World Cup workload.*

*b) The simulated experiments of consolidating two heterogeneous workloads*

The period of a typical workload trace is often weeks, or even months. To evaluate a system, many key factors have effects on experiment results, and we need do many times of time consuming experiments. So we use the simulation method to speedup experiments. We speed up the submission and completion of jobs by a factor of 100. This speedup allows two weeks trace to complete in about three hours.

*c) The simulated clusters*

The workload traces are obtained from platforms with different configurations. For example, NASA iPSC is obtained from the cluster system with each node composed of one CPU; SDSC BLUE is obtained from the cluster system with each node composed of eight CPU; The World Cup resource consumption trace is obtained from the four-core Intel(R) Xeon(R) platform; In the rest of experiments, *our simulated cluster system is modeled after the NASA iPSC cluster, comprising only single-CPU nodes.* So we divide the workload trace of SDSC BLUD by 8.

*d) Synthetic heterogeneous workloads*

To the best of our knowledge, the real traces of parallel batch jobs and Web service on the same platform are not available. However, the focus of our paper is to simulate the case of consolidating *two heterogeneous workloads with different peak resource demands* on a Cloud site. So in our experiments, on a basis of workload traces introduced in Section 6.2, we scale two heterogeneous workload traces with different *constant* factors. We propose a tuple of ($PRC_{PBJ}$, $PRC_{WS}$) to represent two *synthetic heterogeneous workload traces*, where $PRC_{PBJ}$ is the peak resource demand of parallel batch job trace and $PRC_{WS}$ is the peak resource demand of Web service trace. For example, *a tuple of (100, 60)* that is scaled on a basis of SDSC BLUE and World Cup traces means that we respectively scale SDSC BLUE and World Cup traces with two different constant factors, and on the same simulated cluster system, the peak resource demand of SDSC BLUE and World Cup is respectively 100 nodes and 60 nodes.

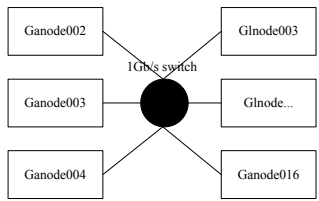

**Fig. 7. The testbed.**

*f) The testbed*

Shown in Fig.7, the testbed includes two types of nodes, nodes with the name starting with glnode and nodes with the name starting with ganode. The nodes of glnode have the same configuration, and each node has 2G memory and two CPUs. Each CPU of the node of glnode has four cores, Intel(R) Xeon(R) (2.00GHz). The OS is 64-bit Linux with kernel of 2.6.18-xen. The nodes of "ganode" have same configuration, and each node has 1G memory and 2 CPUs, AMD Optero242 (1.6GHz). The OS is 64-bit Linux with kernel version of 2.6.5-7.97-smp. All nodes are connected with a 1 Gb/s switch.

## 6.4. The real experiments of World Cup workload

On each node of glnode, we deploy eight XEN [39] virtual machines. For each XEN virtual machine, one core and 256M memory is allocated, and the guest operating system is 64-bit CentOS with kernel version of 2.6.18-XEN.

On the testbed, we deploy a WS TRE shown in Fig.6. In the experiments, *the load balancer* is LVS [27] with direct route mode [18].

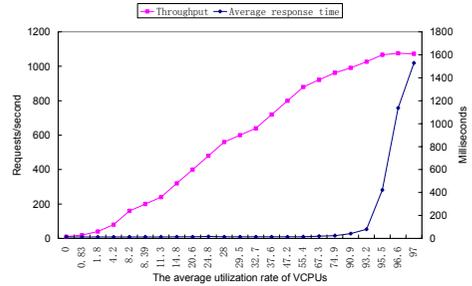

**Fig.8-1. Relationship between actual throughput and average utilization rate of VCPUs on the testbed of 16 virtual machines.**

*Each agent* and *each monitor* are deployed on each virtual machine. LVS and other services are deployed on ganode004, since all of them have light load. We choose *the least-connection scheduling policy* [18] to distribute requests. We choose *httperf* [17] as load generator and open source application ZAP! [19] as the target Web service. The versions of httperf, LVS and ZAP! are respectively 0.9.0, 1.24 and 1.4.5. Two httperf instances are deployed on ganode002 and ganode003.

The Web workload trace is obtained from *the World Cup workload trace* [2] *with a scaling factor of 2.22*. The experiments include *two steps*. First, we decide *the instance adjustment policy*; secondly, we obtain the resource consumption trace.

In the first step, we deploy PhoenixCloud *with the instance adjustment policy disabled*. For this configuration, *the WS manager* will not adjust the number of Web service instances. On the testbed of 16 virtual machines, 16 instances of ZAP! are deployed with each instance deployed on each virtual machine. When *httperf* generates different scale of load, we record the actual throughput, the average response time and the average utilization rate of CPU cores. Since one CPU core is allocated to one virtual machine, for virtual machine, the number of VCPUs is number of CPU cores. So the average utilization rate of each CPU core is also the average utilization rate of VCPUs. Fig.8-1 shows the relationship between the actual throughput and average utilization rate of VCPUs. From Fig.8-1, we observe that when the average utilization rate of VCPUs is below 80%, the average response time of requests is less than 50 milliseconds. However, when the average utilization rate of VCPUs increases to 97%, the average response time of requests dramatically increase to 1528 milliseconds.

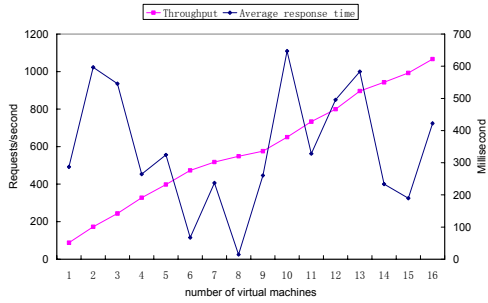

**Fig.8-2. Throughput and average response time V.S. number of virtual machines.**

Based on the above observation, we choose the average utilization rate of VCPUs as the criterion of adjusting the number of instances of ZAP!, and set 80% as the threshold value.

For ZAP!, we specify *the instance adjustment policy* as follows: the initial number of Web service instances is two. If the average utilization rate of VCPUs consumed by all instances of Web service exceeds 80% in the past 20 seconds, *the WS manager* will add one instance. If the average utilization rate of VCPUs, consumed by the current instances of Web service, is lower than (80% * (n-1)/ n) in the past 20 seconds, and *n* is the number of current instances, *the WS manager* will decrease one instance.

In the second step, we deploy PhoenixCloud with the above instance adjustment policy enabled. *The WS manager* adjusts the number of Web service instances according to *the instance adjustment policy*. In the experiments, we also record the relationship between the actual throughputs, the average response time and the number of virtual machine.

From Fig.8-2, we observe that for different number of VMs, the average response time is below 700 milliseconds and the throughput increases linearly with the number of VM increases. This indicates that the instance adjust policy is appropriate, may not optimal.

With the above policies, we obtain the resource consumption trace in two weeks. Fig.9 shows the World Cup resources consumption trace, of which the peak resources demand is 64 VM.

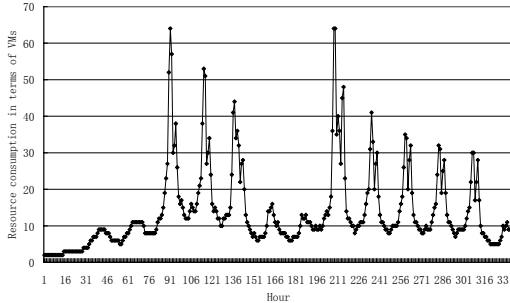

**Fig. 9. The World Cup resource trace in two weeks.**

In the following simulation experiments, if PRC$_{WS}$ is the same in different (PRC$_{PBJ}$, PRC$_{WS}$) tuples, we use the same World Cup resource trace as the input of Web services in DCS, PhoenixCloud and EC2+RightScale.

**6.5 Simulation Experiments of DCS and PhoenixCloud**

In this section, we compare DCS and PhoenixCloud *in the first Cloud scenario* that a resource provider owns the fixed resources that satisfy the resource requests of two REs for heterogeneous workloads

### 6.5.1 The simulated systems
*a) The simulated DCS system*

Since the configuration of a DCS is decided by the peak resource demand of a workload, for a workload tuple (PRC$_{PBJ}$, PRC$_{WS}$) we presume that *the configuration size* of the simulated cluster system is the sum of PRC$_{PBJ}$ and PRC$_{WS}$, which is also the smallest valid configuration size. The left figure in Fig.10 shows the simulated DCS. Resources are statically allocated to two REs: PRC$_{PBJ}$ size for *a PBJ RE* and PRC$_{WS}$ size for *a WS RE*. *The job simulator* is used to simulate the process of submitting job.

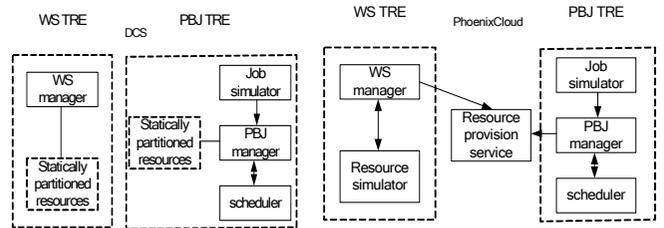

Fig. 10. Simulated DCS and PhoenixCloud systems.
*b) The simulated PhoenixCloud system*

For a workload tuple (PRC$_{PBJ}$, PRC$_{WS}$), in PhoenixCloud, we presume that *the bound of the configuration size of* the simulated cluster system is the sum of PRC$_{PBJ}$ and PRC$_{WS}$. However, *the configuration size* of the simulated cluster may decrease to a lower value.

In comparison with the real PhoenixCloud system in Fig.6, our emulated PhoenixCloud in Fig.10 keeps *the resource provision service*, *the PBJ manager*, *the WS manager* and *the scheduler*, while other services are removed. For *a WS TRE*, *the resource simulator* simulates the varying resources consumption and drives *the WS manager* to request or release resources from or to *the resource provision service*.

### 6.5.2 Experiment configurations
(1) The resource coordination and management policy. For DCS, resources are statically allocated to a RE. PhoenixCloud adopts *the FB policy*.

(2) The scheduling policy. DCS and PhoenixCloud adopt *the same first-fit scheduling policy* for parallel batch jobs. *The first-fit scheduling policy* scans all the queued jobs in the order of job arrival and chooses the first job, whose resources requirement can be met by the system, to execute.

### 6.5.3. Simulation Experiment Results
Table 1-1 and Table 1-2 respectively summarize the experiment results for NASA iPSC+World Cup, of which the tuple of peak resource demands (PRC$_{PBJ}$, PRC$_{WS}$) is *(128, 128),* and SDSC BLUE+World Cup, of which the tuple of peak resource demands (PRC$_{PBJ}$, PRC$_{WS}$) is *(144, 128)*.

TABLE 1-1

TABLE 1-1
METRICS OF DCS AND PHOENIXCLOUD FOR NASA iPSC+WORLD CUP

| System (configuration size) | number of completed jobs | average execution time (seconds) | average turnaround time (seconds) |
|---|---|---|---|
| DCS (256) | 2603 | 573 | 578 |
| PhoenixCloud (128) | 2549 | 520 | 839 |
| PhoenixCloud (152) | 2603 | 573 | 795 |
| PhoenixCloud (217) | 2603 | 573 | 579 |
| PhoenixCloud (256) | 2603 | 573 | 578 |

TABLE 1-2
METRICS OF DCS AND PHOENIXCLOUD FOR SDSC BLUE+WORLD CUP.

| System (configuration size) | number of completed jobs | average execution time (seconds) | Average turnaround time (seconds) |
|---|---|---|---|
| DCS (272) | 2649 | 1975 | 2667 |
| PhoenixCloud (144) | 2591 | 1983 | 7976 |
| PhoenixCloud (163) | 2648 | 1976 | 3438 |
| PhoenixCloud (190) | 2652 | 1977 | 2523 |
| PhoenixCloud (272) | 2657 | 1975 | 2051 |

From Table 1-1 and Table 1-2, we can observe two facts: first, using the FB policy in PhoenixCloud, when the *configuration size* of the simulated cluster is no more than 85% of that of DCS, the throughput of PhoenixCloud is higher than that of DCS (BLUE+WorldCup) or same like that of DCS(iPSC +WorldCup); at the same time, the average turnaround time of PhoenixCloud is better than that of DCS (BLUE+World Cup) or close to that of DCS(iPSC +WorldCup).

Second, when the throughput is almost same like that of DCS with small amount delay of the average turnaround time (maximally by 38%), the *configuration size* of the simulated cluster system can be decreased by about 40% for two groups of heterogeneous workloads.

This is because: (a) for both Web service and parallel batch jobs, the ratios of peak load to normal load are high. However, the peak loads of two traces have different timing; (b) when Web service has a short spike, the FB policy will kill running jobs with the smallest resource demands, so we can decrease the configuration size of cluster system, but at the same time increase the average turnaround time.

When $PRC_{PBJ}$ is the same, Table 2-1 and Table 2-2 show the effect of different ratios of $PRC_{WS}$ to $PRC_{PBJ}$ on the performance metrics of PhoenixCloud.

TABLE 2-1.
Metrics of PhoenixCloud for iPSC+WorldCup.

| ($PRC_{PBJ}$, $PRC_{WS}$), configuration size | Saved resources (%) with respect to DCS | number of completed jobs | average turnaround time (seconds) |
|---|---|---|---|
| (128,64), 128 | 33% | 2549 | 575 |
| (128,128),128 | 50% | 2549 | 839 |
| (128,256),256 | 33% | 2603 | 676 |

TABLE 2-2.
Metrics of PhoenixCloud for BLUE+WorldCup.

| ($PRC_{PBJ}$, $PRC_{WS}$), configuration size | saved resources (%) with respect to DCS | number of completed jobs | average turnaround time (seconds) |
|---|---|---|---|
| (144,64), 144 | 31% | 2636 | 3343 |
| (144,128),144 | 47% | 2591 | 7976 |
| (144,256),256 | 36% | 2657 | 2609 |

From Table 2-1 and Table 2-2, we can observe that when two peak resource demands in ($PRC_{PBJ}$, $PRC_{WS}$) are close, *the percent of saved resources*, which is obtained with the smallest *configuration size* of cluster, outperforms other cases. This is because when we consolidate two heterogeneous workloads, the *configuration size* of PhoenixCloud must be greater than the maximum value of two peak resource demands. For parallel batch jobs, if the *configuration size* of cluster is less than the resource demand of the biggest job, the biggest job can not run. For Web service, if the *configuration size* of cluster is less than the peak resource demand, overload will happen.

### 6.6 Simulation Experiments of EC2+RightScale and PhoenixCloud

In this section, we compare the performance of PhoenixCloud and EC2+RightScale *in the second Cloud scenario*. We presume that the simulated cluster system has abundant resources with respect to resource requests of two heterogeneous workloads in both two systems.

#### 6.6.1 The simulated systems

*a) The simulated EC2+RightScale system*

Because RightScale provides the same scalable management for Web service as PhoenixCloud, we just use the same resource consumption trace for Web service in two systems. For parallel batch jobs, in EC2, end users simultaneously request resources needed by parallel batch jobs, and the submitted jobs will run immediately, so there is no need for *the scheduler*. Fig. 11 shows the simulated architecture of EC2+RightScale.

*b) The simulated PhoenixCloud*

The simulated PhoenixCloud is same as that shown in Fig.10 with *the FLB-NUB policy*.

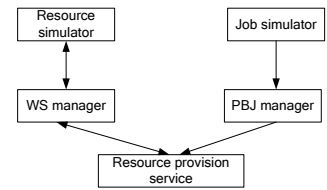

**Fig.11. The simulated system of EC2+RightScale.**

#### 6.6.2 Experiment configurations

(1) The resource coordination policy. For PhoenixCloud, we adopt the FLB-NUB policy. For EC2 + RightScale, There is no resource coordination between two REs.

(2) The scheduling policy of parallel batch jobs. PhoenixCloud adopt the first-fit scheduling policy. EC2 needs no scheduling policy, since it is each end user that is responsible for running parallel batch jobs.

(3) The resource management policy. For both systems, there is *a time unit of leasing resources*. We presume that *the lease term of a resource is a time unit of leasing resource times a positive integer*. In the EC2+RightScale solution, for parallel batch jobs, each end user is responsible for manually managing resources on EC2 system, and we presume that *a user only releases resource at the end of each time unit of leasing resources if a job runs over*. This is because: a) EC2 charges the usage of resources in terms of a *time unit of leasing resources* (an hour); b) It is difficult for end user to predict the completed time of jobs, and releasing resources to resource provider on time is almost impossible. PhoenixCloud adopt the FLB-NUB policy

### 6.6.3 Experiment Results

Before reporting experiment results, we pick the following parameters as the baseline configuration of PhoenixCloud for comparison, and detailed parameter analysis will be deferred to Section 6.6.4.

Through comparisons with large amount of experiments, we set the baseline parameters in PhoenixCloud:[B25/U1.2/V0.2/G0.5] for iPSC+World Cup and [B27/U1.2/V0.2/G0.5] for SDSC+WorldCup, where [B25/U1.2/V0.2/G0.5] indicates that *the size of coordinated resources* (which is represented as *B*) is 25 nodes, *the threshold ratio of requesting resources* (which is represented as U) is 1.2; V0.2 indicates that *the threshold ratio of releasing resources* (which is presented as V) is 0.2; G0.5 indicates that *the elastic factor of releasing resources* (which is represented as G) is 0.5. In both two systems, *the time unit of leasing resources* (which is represented as L) is 60 minute.

Table 3-1 and Table 3-2 respectively summarize the experiment results for iPSC+WorldCup, of which ($PRC_{PBJ}$, $PRC_{WS}$) is *(128, 128)*, and BLUE+WorldCup traces, of which is ($PRC_{PBJ}$, $PRC_{WS}$) is *(144, 128)*. From two tables, we can observe two facts: (1) The total resource consumption of PhoenixCloud is less than that of EC2+RightScale (maximally by 28% and minimally by 14%) with delay of average turnaround time per jobs (maximally by 44% and minimally by 35%); (2) PhoenixCloud decreases peak resource consumption maximally to 31% with respect to that of EC2+RightScale. This is because that PhoenixCloud only requests resources on the condition that *the threshold ratio of requesting resources* is exceeded, or else jobs will be queued, so PhoenixCloud decreases peak resource consumption and total resource consumption, and increases the average turnaround time.

TABLE 3-1
METRICS OF EC2+RIGHTSCALE AND PHOENIXCLOUD FOR iPSC +WORLDCUP.

| system | number of completed jobs | average turnaround time | Peak resource consumption | Total resource consumption |
|---|---|---|---|---|
| EC2+RightScale | 2603 | 573 seconds | 1319 nodes | 63336 node*hour |
| PhoenixCloud | 2603 | 826 seconds | 412 nodes | 45803 node*hour |

TABLE 3-2
METRICS OF EC2+RIGHTSCALE AND PHOENIXCLOUD FOR SDSC+WORLDCUP.

| system | number of completed jobs | Average turnaround time | Peak resource consumption | Total resource consumption |
|---|---|---|---|---|
| EC2+RightScale | 2657 | 1975 seconds | 834 nodes | 45056 node* hour |
| PhoenixCloud | 2656 | 2669 seconds | 468 nodes | 38623 node* hour |

When $PRC_{PBJ}$ is the same, Table 4-1 and Table 4-2 show the effect of different ratios of $PRC_{WS}$ to $PRC_{PBJ}$ on the performance metrics of PhoenixCloud. Due to the space limitation, we constrain most of our discussion to the configuration of BR0.1_U1.2_V0.2_G0.5_L60, *where BR0.1 indicates the ratio of the size of the coordinated resources of PhoenixCloud to the sum of $PRC_{WS}$ and $PRC_{PBJ}$ is 0.1*

From Table 4, we can observe that when the ratio of $PRC_{WS}$ to $PRC_{PBJ}$ increases, the percent of saved resources (%) increases, which is obtained against the sum of $PRC_{WS}$ and $PRC_{PBJ}$. This observation is different from that of the FB policy in Section 6.5.3. This is because in the FLB-NUB policy, resources can be dynamically requested beyond *the lower bound*; while in the FB policy, the resources only can be dynamically requested within the limit of *the lower bound*.

TABLE 4-1.
Metrics of PhoenixCloud for iPSC +WorldCup.

| ($PRC_{PBJ}$, $PRC_{WS}$) | number of completed jobs | Average execution time(seconds) | Average turnaround time(seconds) | Saved resources (%) |
|---|---|---|---|---|
| (128,64) | 2603 | 573 | 839 | 38.3% |
| (128, 128) | 2603 | 573 | 826 | 46.8% |
| (128,256) | 2603 | 573 | 839 | 58.5% |

TABLE 4-2.
Metrics of PhoenixCloud for BLUE+WorldCUP.

| ($PRC_{PBJ}$, $PRC_{WS}$) | number of completed jobs | Average execution time(seconds) | Average turnaround time(seconds) | Saved resources (%) |
|---|---|---|---|---|
| (144,64) | 2654 | 1974 | 2682 | 52.0% |
| (144, 128) | 2656 | 1975 | 2669 | 57.7% |
| (144,256) | 2654 | 1974 | 2761 | 64.5% |

### 6.6.4 Parameter Analysis

Because of space limitation, we are unable to present the data for the effect of all parameters; instead, we constrain most of our discussion to the configuration that one parameter varies while the other parameters keep the same as those of the baseline configuration in Section 6.6.3, which are representative of the trends that we observe across all cases.

**The effect of the size of coordinated resources (B)**. To save space, in PhoenixCloud we tune B, while other parameters are *[U1.2/V0.2/G0.5/L60]*. Fig.12 and Fig.13 shows the effect of different *B for two groups of heterogeneous workloads*. In the rest of this section, tuples of ($PRC_{PBJ}$, $PRC_{WS}$) of iPSC+WorldCup and BLUE+WorldCUP are respectively (128, 128) and (144, 128).

From Fig.12 and Fig.13, we have the following observations:

1) With the increase of *B*, the total resource consumption increases, while the average turnaround time decreases. This is because resources under the *lower bound* are only allocated to PBJ TRE and WS TRE, hence idle resources will also increase when *B* increases for the same workload; at the same time, with the increase of *B*, more resources will be provisioned to PBJ TRE, so the average turnaround time per jobs decreases.

2) The change of *B* has small effect on the number of completed jobs. This is because PhoenixCloud can dynamically request resources when the threshold ratio of requesting resource is triggered.

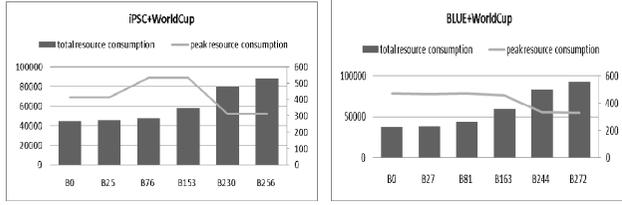

*Fig.12. Peak and total resource consumptions V.S. different* **B.**

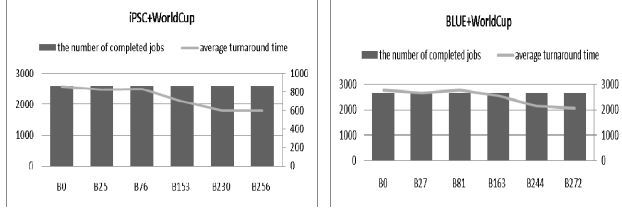

*Fig.13. The number of completed jobs and average turnaround time V.S. different* **B.**

**The effects of the threshold ratios of requesting resources and releasing resources (V and V) and the elastic factor of releasing resource (G).** To save space, in PhoenixCloud we tune one of *U, V, G*, while other parameters are *[B25/U1.2/V0.2/G0.5/L60 ]* for iPSC +WorldCup and *[B27/U1.2/V0.2/G0.5/L60 ]* for BLUE +WorldCUP. Fig.14 and Fig.15 show the effect of different parameters.

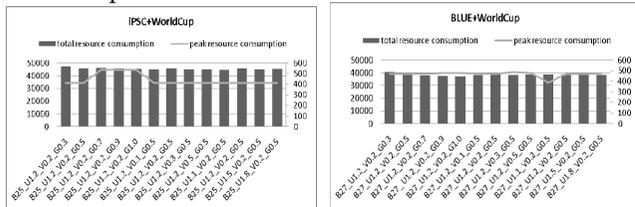

*Fig.14. peak and total resource consumptions V.S. different G, V, U.*

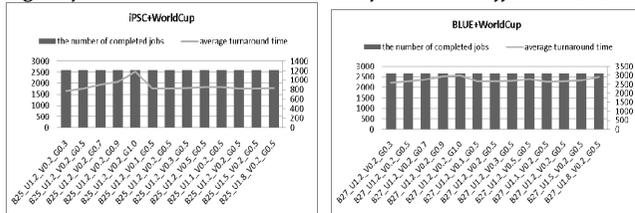

*Fig.15 the number of completed jobs and average turnaround time V.S. different G, V, U.*

From Fig.14 and Fig.15, we have the following observations:

1) *U, V, G* have small effect on *the total resource consumption* and *the number of completed jobs* when *B* is fixed.

2) *G* is proportional to *the average turnaround time* when *B* is fixed. This is because larger elastic factor of releasing resources will result in less idle resources when new jobs are submitted. *U* and *V* have small effect on *the average turnaround time*.

**The effects of the time unit of leasing resources.** We respectively set *the time unit of leasing resources L* as 15/30/60/120/240 minutes, while other parameters are *[B25/U1.2/V0.2/G0.5]* for NASA iPSC workload and *[B27/U1.2/V0.2/G0.5]* for SDSC BLUE workload. In Fig.16, *iPSC-15* implies that *L* is 15 minutes and workload is iPSC.

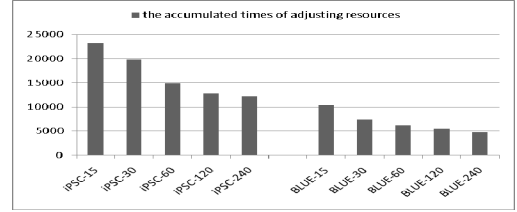

**Fig.16. management overhead V.S. different time unit of leasing resources.**

From Fig. 16, we have the following observation:

1) The management overhead is inversely proportional to *L*. This is because when the time unit of leasing resources is less, the service provider requests resources more frequently.

Taking it into account resources are charged at the granularity of a time unit of leasing resources, we make a tradeoff and select *L* as 60 minutes in PhoenixCloud and EC2+RightScale. In fact, in EC2 system, resources are also charged at the granularity of one hour.

**Implications of Analysis.** Based on the above analysis, we have the following suggestions in choosing factors for two coordinated REs for Web service and parallel batch jobs: since the increase of *B* will also result in the increase of total resource consumption, we suggest selecting a low value for *B*: about 10% of the sum of $PRC_{PBJ}$ and $PRC_{WS}$. Increasing the elastic factor of releasing resource *G* will result in the delay of the average turnaround time**.** Our experiments show *0.5* makes a good compromise. According to our experiments, when *U* is greater than 1.0 and less than 2.0, it has small effects on the metrics in our experiments; when *V* is greater than 0.1 and less than 0.5, it has small effects on the metrics in our experiments. So we suggest service providers choose baseline configuration in Section 6.6.3. for *U, V, G*.

## 6.7 Discussions

Our experiments show that a service provider has three choices in consolidating heterogeneous workloads:

1) If resorting to a private Cloud with the fixed size, he should choose PhoenixCloud with the FB policy. With this solution, the configuration size is smallest with respect to other three solutions. However, this solution increases both the average execution time and the average turnaround time, since jobs may be killed to reallocate resources to web services.

In a public Cloud scenario,

2) If paying high attention to the average turnaround time per jobs, he should choose EC2+RightScale solution. However, this solution will result in higher peak resource consumption, which is several times (two or three in our experiments) of that of PhoenixCloud, and larger total resource consumption.

3) If making a tradeoff among the resource consumption and the average turnaround time of jobs, he should choose PhoenixCloud with the FLB-NUB policy. With this solution, the total and peak resource consumptions of PhoenixCloud are smaller that that of EC2+RightScale, while the average turnaround time is larger than that of EC2+RightScale *with small delay*.

## 7. CONCLUSIONS

In this paper, we presented a RE agreement that express diverse RE requirements and build an innovative system PhoenixCloud to enable creating REs on demand according to RE agreements. For two typical heterogeneous workloads: *Web services* and *parallel batch jobs*, we proposed two coordinated resource provisioning solutions in two different Cloud scenarios.

For three typical workload traces: SDSC BLUE, NASA iPSC and World Cup, our experiments showed that: a) in the first Cloud scenario, when the throughput is almost same like that of a DCS, our solution decreases the configuration size of cluster by about 40%; b) in the second Cloud scenario, our solution decreases not only the total resource consumption, but also the peak resource consumption maximally to 31% with respect to that of EC2 + RightScale solution.

heterogeneous workloads of large organizations on cloud computing platforms. The first workshop of Cloud computing and its application (CCA 08). Chicago. 2008.